\documentclass[amsmath,twocolumn,prx,footinbib,longbibliography]{revtex4-2} %nofootinbib
\usepackage{amsthm,amssymb,amsfonts,graphicx,verbatim,bm} % The basics
\usepackage[dvipsnames]{xcolor}
\usepackage{hyperref} % URL links
\usepackage[utf8]{inputenc} % encoder
\usepackage{siunitx} % nice units
\usepackage{tikz} % drawing in-line
\usetikzlibrary{arrows,decorations.pathmorphing,backgrounds,positioning,fit,matrix}

\let\oldmarginpar\marginpar
\renewcommand\marginpar[1]{\-\oldmarginpar[\raggedleft\footnotesize #1]%
{\raggedright\footnotesize #1}}
\renewcommand{\eqref}[1]{(\ref{#1})}
\newcommand{\diag}{\mathrm{diag}}
\newcommand{\Tr}{\mathrm{Tr}}
\newcommand{\ket}[1]{|#1\rangle}
\newcommand{\bra}[1]{\langle #1|}

\begin{document}

\title{Dimer states of Rydberg atoms on the Kagome lattice as resources for universal measurement-based quantum computation}
\author{Valentin Cr\'epel}
\affiliation{Department of Physics, Massachusetts Institute of Technology, 77 Massachusetts Avenue, Cambridge, MA, USA}

\begin{abstract}
We show that the quantum dimer state on the Kagome lattice, which was recently realized with high fidelity in a Rydberg quantum simulator [Semeghini \textit{et al.}, Science 374, 6572 (2021)], offers a sufficient resource for universal measurement-based quantum computations. In particular, we provide an efficient encoding of logical qubits in this state, and give explicit measurement sequences that implement a universal set of gates on these qubits. Since the building blocks of the proposed measurements have already been experimentally implemented, our work highlights one possible path towards promoting Rydberg simulators to universal quantum computers relying on the measurement-based model of quantum computation with currently existing technology. 
\end{abstract}

\maketitle

\section{Introduction}

Quantum simulators are controlled quantum devices that can be used to simulate other quantum systems, providing the means to study a class of problems currently intractable on classical computers~\cite{buluta2009quantum,georgescu2014quantum,altman2021quantum}. Being special purpose devices, they are often contrasted with general-purpose universal quantum computers~\cite{brown2010using,johnson2014quantum}, which are -- in principle -- capable of solving any quantum problems, including those addressed by simulators. This apparent, but loosely stated, inclusion relation is however misleading, as it is in fact an equivalence.

Indeed, certain quantum entangled states, once simulated, provide sufficient resources for the realization of any quantum algorithm using the model of measurement-based quantum computation (MBQC)~\cite{gottesman1999quantum,raussendorf2001one,nielsen2003quantum,leung2004quantum}, wherein measurements on individual constituents of the entangled state drive the algorithm. Within the MBQC paradigm, quantum simulators realizing those computationally universal states (CUSs) are therefore equivalent to universal quantum computers~\cite{jozsa2006introduction,briegel2009measurement}.

The standard example of such CUS, and the first to have been identified, is the cluster state on the square lattice~\cite{raussendorf2003measurement,nielsen2006cluster}. Since then, several other two dimensional states -- such as graph states~\cite{van2006universal,gross2007novel}, the tricluster state~\cite{chen2009gapped}, modified toric code states~\cite{gross2007measurement}, and Affleck-Kennedy-Lieb-Tasaki states~\cite{wei2011affleck,miyake2011quantum} -- have been diagnosed as CUSs. Intriguingly, all these candidates for universal MBQC share a common feature, they are two-dimensional symmetry-protected topologically ordered states (SPTOSs)~\cite{else2012symmetry,nautrup2015symmetry,wei2017universal,stephen2017computational,raussendorf2019computationally}. The coincidence between CUS and SPTOS was further grounded numerically in certain models~\cite{darmawan2012measurement,huang2016emergence}. Using this phenomenological observation, the search for quantum simulators that may achieve universal quantum computation focused on those realizing two-dimensional SPTOS.

Recently, a two-dimensional state with $\mathbb{Z}_2$ topological order has been realized in Rydberg arrays~\cite{semeghini2021probing}, one of the most promising quantum simulator for the study of many-body systems with short-range interactions~\cite{weimer2010rydberg,browaeys2016interacting,morgado2021quantum}. Extensive numerical studies~\cite{verresen2021prediction,verresen2021efficiently} have shown that this SPTOS is described with very high accuracy by a quantum dimer state on the Kagome lattice; the overlap between these two states reaching as high as 99$\%$ for 48 atoms under realistic experimental conditions~\cite{giudici2022dynamical}. The Kagome dimer state has been extensively studied in the literature~\cite{zeng1995quantum,poilblanc2010effective,hao2014destroying} and is known to host a $\mathbb{Z}_2$ topological order~\cite{misguich2002quantum,nikolic2003physics,singh2008triplet} closely related to Kitaev's toric code~\cite{kitaev2006anyons,bravyi2007measurement}, which rationalizes the experimental results of Ref.~\cite{semeghini2021probing}.

The observed relation between CUS and SPTOS naturally raises the question whether the $\mathbb{Z}_2$ topological state of Ref.~\cite{semeghini2021probing} exhibits universal computational power within the MBQC paradigm.

In this article, we answer to this question positively. To achieve this goal, we first model the Kagome dimer state as a pair-entangled projected pair state (PEPS)~\cite{cirac2021matrix}, which also provides an intuitive encoding of logical qubit on this state. We then provide explicit realistic measurement sequences implementing a universal set of quantum gates on these qubits. Importantly, we have designed these measurements to only use time evolution and fluorescence measurements akin to those already realized in current Rydberg arrays experiments. Our realistic MBQC scheme hints at one possible route for achieving universal quantum computations using Rydberg arrays. However, this route will necessarily involve many technological advances, in particular in the quality of state preparation, as only quantum code consisting of about ten gates can be successfully run under current experimental conditions.

This small algorithmic depth calls for an extensive search of UCS in Rydberg arrays, tracking those offering the simplest experimental sequences for the implementation of a universal gate set. Experimental imperfections weakening the reliability of MBQC, such as non-adiabaticity or dephasing due to long-range interactions, should also be identified and listed, together with ways to correct for them, \textit{i.e.} through pulse-engineering. Their careful analysis goes beyond the scope of this paper and will be the focus of future works which will use more appropriate analytical and numerical techniques.

\section{Required experimental resources}

Let us first describe the SPTOS realized in Ref.~\cite{semeghini2021probing} and characterize the types of measurement available in typical Rydberg array experiments. These two types of resources will later serve as building blocks for our MBQC scheme.

\subsection{$\mathbb{Z}_2$ SPTOS in a Rydberg simulator}

\begin{figure}
\centering 
\includegraphics[width=\columnwidth]{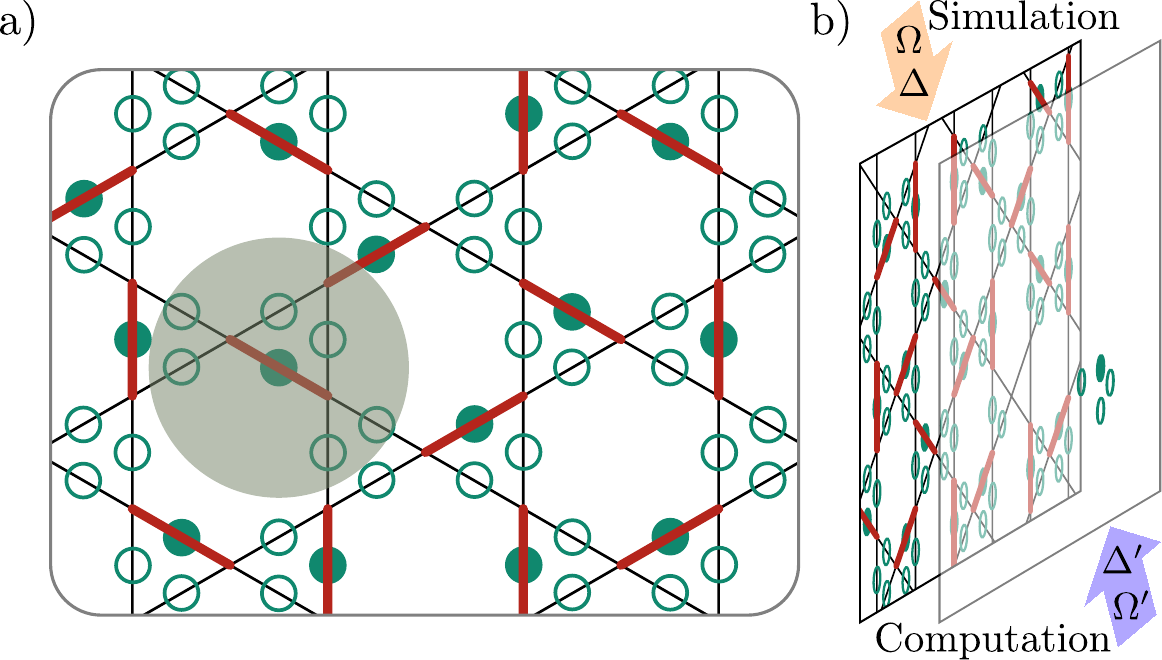}
\caption{a) State consistent with the blockade condition (grey circle) at filling $\langle n \rangle = 1/4$. Atoms, either in $\ket{g}$ (empty) and $\ket{e}$ (filled circles), are placed on the links of a Kagome lattice, \textit{i.e.} on a Ruby lattice. The equivalence with dimer configurations is made explicit by coloring the links of the Kagome lattice on which Rydberg excitations lie. b) Physical decoupling of the simulation and computation (or measurement) plane allow for pulses with arbitrary driving parameters $(\Omega',\Delta')$.}
\label{fig_RydbergToDimer}
\end{figure}

The Rydberg array of Ref.~\cite{semeghini2021probing} consists of atoms individually trapped in optical tweezers~\cite{barredo2016atom,scholl2021quantum,ebadi2021quantum}, and positioned on a Ruby lattice, \textit{i.e.} on the links of a Kagome lattice (see Fig.~\ref{fig_RydbergToDimer}a). These atoms are initially prepared in an electronic ground state $\ket{g}$, which is laser-coupled to a Rydberg state $\ket{e}$ with a Rabi frequency $\Omega$ and a detuning $\Delta$, leading to the on-site Hamiltonian  
\begin{equation} \label{eq_totalHamiltonian} 
\mathcal{H} = \frac{1}{2} \left(\Omega \ket{e} \bra{g} + \Omega^* \ket{g} \bra{e} \right) - \Delta n , \quad n = \ket{e} \bra{e} .
\end{equation}
Due to their high polarizability, atoms in the Rydberg states strongly interact through the strong van der Waals interaction $V(d) = C_6 / d^6$, with $d$ the interatomic distance~\cite{browaeys2016interacting}. As a result, all atoms within a distance $r<R_b = (C_6/\Omega)^{1/6}$ of an atom in $\ket{e}$ are brought far off resonance from the laser field. This effect is well captured by the so-called blockade radius model, in which the system's dynamics is approximated by Eq.~\ref{eq_totalHamiltonian} projected to the subspace where no pairs of excitations distant by less than $R_b$ are present.

In Ref.~\cite{semeghini2021probing}, the Rabi frequency and detuning have been set such that $2a < R_b < \sqrt{7} a$, $a$ being the Ruby lattice constant, and $\langle n \rangle = 1/4$. The first condition prevents adjacent links of the Kagome lattice to simultaneously host an atom in $\ket{e}$, while the second requires that at least one of the four links surrounding each vertex be in the Rydberg state. As sketched in Fig.~\ref{fig_RydbergToDimer}a, these conditions restrict many-body configurations to dimer coverings of the Kagome lattice, where the position of dimers is given by the excited atoms~\cite{samajdar2021quantum}. Including the effects of Eq.~\ref{eq_totalHamiltonian} in this subspace produces terms that favor the equal superposition of all those dimer coverings~\cite{verresen2021prediction}. Altogether, these arguments outline intuitive reasons explaining why the dimer state $\ket{\Phi}$, \textit{i.e.} the equal superposition of all existing dimer coverings, can be realized by the Rydberg blockade Hamiltonian. This state has been thoroughly studied in the literature~\cite{zeng1995quantum,poilblanc2010effective,hao2014destroying} and is known to host $\mathbb{Z}_2$ topological order~\cite{misguich2002quantum,nikolic2003physics,singh2008triplet}.

Experimentally, the SPTOS of Ref.~\cite{semeghini2021probing} was realized dynamically from the state with no Rydberg excitation by ramping up the parameters $\Omega$ and $\Delta$ to values compatible with  $2a < R_b < \sqrt{7} a$ and $\langle n \rangle = 1/4$. Because of this out of equilibrium preparation, the Rydberg array does not necessarily probe the ground state properties highlighted above. Nevertheless, considering realistic experimental conditions, extensive numerical calculations have estimated that the overlap between the dynamically prepared and dimer states reach as high as 99$\%$ for 48 atoms~\cite{giudici2022dynamical}. While reproducing these calculations fall beyond the scope of the present work, the very high overlap obtained in Ref.~\cite{giudici2022dynamical} allows us to confidently identify the SPTOS of Ref.~\cite{semeghini2021probing} with the dimer state $\ket{\Phi}$ on the Kagome lattice.

The first goal of this paper is to explore the computational power of $\ket{\Phi}$ within MBQC, and prove that it is a CUS. Our second aim is to show that universal MBQC is within experimental reach. The measurements sequences proposed in this article should therefore remain realistically implementable in current Rydberg array experiments.

\subsection{Available measurements} 

To meet this last criterion, we shall only rely on fluorescence measurement of the Rydberg excitation number at each site, potentially preceded by a 'pulse'. A pulse is defined as the free evolution of a few atom cluster decoupled from the rest of the lattice under the Rydberg blockade model Hamiltonian. Note that pulses might require different laser parameters ($\Omega'$, $\Delta'$) than those used to stabilize the dimer state. One possible setup allowing for this discrepancy is depicted in Fig.~\ref{fig_RydbergToDimer}b: atomic clusters are moved out of the simulation plane into a computation plane, where a different laser is shone to produce the desired pulse and where the final fluorescence measurement takes place~\cite{barredo2018synthetic}.

In the next section, we will combine three elementary pulses to form a universal set of gate on the encoded logical qubits: (I) The first one involves a two-atom cluster, evolved under Eq.~\ref{eq_totalHamiltonian} for a time $\pi / \Delta$, with the detuning tuned such that $\Delta / \Omega = \sqrt{2/3}$~\footnote{This choice makes the effective Rabi frequency $\tilde{\Omega} = \sqrt{2\Omega^2 + \Delta^2} = 2\Delta$ commensurate with the detuning.}. The Hamiltonian in the two-atom Rydberg blockaded subspace 
$\{
\begin{tikzpicture}[scale=0.3,baseline=-2.7]
\def\x{sqrt(3)}; \draw[thick] (0,0.5) -- ({0.5*\x},0) -- (0,-0.5); 
\end{tikzpicture} , \begin{tikzpicture}[scale=0.3,baseline=-2.7]
\def\x{sqrt(3)}; \draw[thick] (0,0.5) -- ({0.5*\x},0) -- (0,-0.5);  \draw[PineGreen,fill=PineGreen] ({0.25*\x},0.25) circle (0.2); 
\end{tikzpicture} , \begin{tikzpicture}[scale=0.3,baseline=-2.7]
\def\x{sqrt(3)}; \draw[thick] (0,0.5) -- ({0.5*\x},0) -- (0,-0.5);  \draw[PineGreen,fill=PineGreen] ({0.25*\x},-0.25) circle (0.2); 
\end{tikzpicture} \}$
is easily exponentiated to give
\begin{equation} \label{eq_pulse1_evolution}
e^{-i\mathcal{H} \tau} = \begin{bmatrix} -1&0&0 \\0&-q&q^*\\ 0&q^*&-q\end{bmatrix} , \quad  q = \frac{1+i}{2} .  
\end{equation}
The measurement of the Rydberg excitation position $P = \diag(0,-1,1)$ after this time evolution becomes equivalent to the measurement of 
\begin{equation} \label{eq_EffectiveMeasurement}
e^{i\mathcal{H} \tau} P e^{-i\mathcal{H} \tau} = \begin{bmatrix} 0&0&0 \\ 0 & 0&-i\\0&i&0 \end{bmatrix} ,
\end{equation}
on the original lattice, and projects onto the states $\begin{tikzpicture}[scale=0.3,baseline=-2.7]
\def\x{sqrt(3)}; \draw[thick] (0,0.5) -- ({0.5*\x},0) -- (0,-0.5); 
\end{tikzpicture}$, $(\begin{tikzpicture}[scale=0.3,baseline=-2.7]
\def\x{sqrt(3)}; \draw[thick] (0,0.5) -- ({0.5*\x},0) -- (0,-0.5);  \draw[PineGreen,fill=PineGreen] ({0.25*\x},0.25) circle (0.2); 
\end{tikzpicture} + i \begin{tikzpicture}[scale=0.3,baseline=-2.7]
\def\x{sqrt(3)}; \draw[thick] (0,0.5) -- ({0.5*\x},0) -- (0,-0.5);  \draw[PineGreen,fill=PineGreen] ({0.25*\x},-0.25) circle (0.2); 
\end{tikzpicture})$ and $(\begin{tikzpicture}[scale=0.3,baseline=-2.7]
\def\x{sqrt(3)}; \draw[thick] (0,0.5) -- ({0.5*\x},0) -- (0,-0.5);  \draw[PineGreen,fill=PineGreen] ({0.25*\x},0.25) circle (0.2); 
\end{tikzpicture} - i \begin{tikzpicture}[scale=0.3,baseline=-2.7]
\def\x{sqrt(3)}; \draw[thick] (0,0.5) -- ({0.5*\x},0) -- (0,-0.5);  \draw[PineGreen,fill=PineGreen] ({0.25*\x},-0.25) circle (0.2); 
\end{tikzpicture})$ when $p=0,1,-1$ is measured, respectively. (II) The second pulse isolates three atoms within a blockade radius for a time $4 \pi / (3\sqrt{3} \Omega)$ with $\Delta = 0$ before fluorescence measurement. A similar analysis on the blockaded Hilbert space $\{
\begin{tikzpicture}[scale=0.3,baseline=-2.7]
\def\x{sqrt(3)}; \draw[thick] ({0.5*\x},0) -- (0,0.5) -- (0,-0.5) -- ({0.5*\x},0); 
\end{tikzpicture} , \begin{tikzpicture}[scale=0.3,baseline=-2.7]
\def\x{sqrt(3)}; \draw[thick] ({0.5*\x},0) -- (0,0.5) -- (0,-0.5) -- ({0.5*\x},0); \draw[PineGreen,fill=PineGreen] (0,0) circle (0.2); 
\end{tikzpicture} , \begin{tikzpicture}[scale=0.3,baseline=-2.7]
\def\x{sqrt(3)}; \draw[thick] ({0.5*\x},0) -- (0,0.5) -- (0,-0.5) -- ({0.5*\x},0); \draw[PineGreen,fill=PineGreen] ({0.25*\x},0.25) circle (0.2); 
\end{tikzpicture} , \begin{tikzpicture}[scale=0.3,baseline=-2.7]
\def\x{sqrt(3)}; \draw[thick] ({0.5*\x},0) -- (0,0.5) -- (0,-0.5) -- ({0.5*\x},0); \draw[PineGreen,fill=PineGreen] ({0.25*\x},-0.25) circle (0.2); 
\end{tikzpicture} \}$ shows that the state of the atomic cluster is now projected on $\frac{1}{2} (
\begin{tikzpicture}[scale=0.3,baseline=-2.7]
\def\x{sqrt(3)}; \draw[thick] ({0.5*\x},0) -- (0,0.5) -- (0,-0.5) -- ({0.5*\x},0); 
\end{tikzpicture} - p_\uparrow p_\downarrow \begin{tikzpicture}[scale=0.3,baseline=-2.7]
\def\x{sqrt(3)}; \draw[thick] ({0.5*\x},0) -- (0,0.5) -- (0,-0.5) -- ({0.5*\x},0); \draw[PineGreen,fill=PineGreen] (0,0) circle (0.2); 
\end{tikzpicture}  - x p_\uparrow \begin{tikzpicture}[scale=0.3,baseline=-2.7]
\def\x{sqrt(3)}; \draw[thick] ({0.5*\x},0) -- (0,0.5) -- (0,-0.5) -- ({0.5*\x},0); \draw[PineGreen,fill=PineGreen] ({0.25*\x},-0.25) circle (0.2); 
\end{tikzpicture}  - x p_\downarrow \begin{tikzpicture}[scale=0.3,baseline=-2.7]
\def\x{sqrt(3)}; \draw[thick] ({0.5*\x},0) -- (0,0.5) -- (0,-0.5) -- ({0.5*\x},0); \draw[PineGreen,fill=PineGreen] ({0.25*\x},0.25) circle (0.2); 
\end{tikzpicture} )$, where $x$/$p_\downarrow$/$p_\uparrow$ are equal to one when the left/bottom right/top right link of the triangle is in $\ket{e}$, and to minus one otherwise~\cite{verresen2021prediction}. (III) Finally, a far detuned laser field ($\Delta \gg \Omega$) can be applied for a time $t$ to imprint a variable phase shift $\varphi = \Delta t$ on the $\ket{e}$ state of individual atoms. 

In the following paragraphs, we show how MBQC can be implemented by combination of these three measurement pulses on $\ket{\Phi}$. Note that the denomination of measurement is here slightly different meaning than in other studies of MBQC, as the sequences (I-III) are not single qubit measurements, but correlated measurements involving one to three atoms within a blockade radius.

\section{Universal MQBC} \label{sec_MBQCscheme}

Our scheme for MBQC using $\ket{\Phi}$ as a resource follows the principle of computation in correlation space~\cite{gross2007novel,gross2007measurement,cai2009quantum}. More precisely, we write the dimer state as a PEPS, and encode the logical qubits in the auxiliary space of the PEPS tensor. Quantum information processing is performed by pulse measurements that act as special contractions of the tensor's physical space. The residual action of these contracted tensors on the logical qubits executes a desired set of gates.

\subsection{Tensor network description} 

To define our logical qubits and to make the MBQC scheme transparent, we introduce one possible tensor network representation of the dimer state
\begin{equation} \label{eq_TotalStateOnLattice}
\ket{\Phi} =
\begin{tikzpicture}[scale=0.4,baseline=-1]
\def\x{sqrt(3)}	
\foreach \i in {-\x,0,\x} {
\foreach \j in {-2,2} {
\draw[rounded corners] ({(-0.75)*\x+2*\i}, {-0.75+\j}) rectangle ({(0.75)*\x+2*\i}, {0.75+\j}) {};
\draw[thick] ({0.5*\x+2*\i},{-0.5+\j}) -- ({-0.5*\x+2*\i},{0.5+\j}) -- ({-0.5*\x+2*\i},{-0.5+\j}) -- ({0.5*\x+2*\i},{0.5+\j}) -- ({0.5*\x+2*\i},{-0.5+\j}); 
\draw[BrickRed,thick] ({-0.5*\x+2*\i}, {-0.75+\j}) -- ({-0.5*\x+2*\i}, {-1+\j}); \draw[BrickRed,thick] ({0.5*\x+2*\i}, {-0.75+\j}) -- ({0.5*\x+2*\i}, {-1+\j}); \draw[BrickRed,thick] ({-0.5*\x+2*\i}, {0.75+\j}) -- ({-0.5*\x+2*\i}, {1+\j}); \draw[BrickRed,thick] ({0.5*\x+2*\i}, {0.75+\j}) -- ({0.5*\x+2*\i}, {1+\j});
} }
\def\j{0}	
\foreach \i in {-3*\x/2,-\x/2,\x/2,3*\x/2} {
\draw[rounded corners] ({(-0.75)*\x+2*\i}, {-0.75+\j}) rectangle ({(0.75)*\x+2*\i}, {0.75+\j}) {};
\draw[thick] ({0.5*\x+2*\i},{-0.5+\j}) -- ({-0.5*\x+2*\i},{0.5+\j}) -- ({-0.5*\x+2*\i},{-0.5+\j}) -- ({0.5*\x+2*\i},{0.5+\j}) -- ({0.5*\x+2*\i},{-0.5+\j}); 
\draw[BrickRed,thick] ({-0.5*\x+2*\i}, {-0.75+\j}) -- ({-0.5*\x+2*\i}, {-1+\j}); \draw[BrickRed,thick] ({0.5*\x+2*\i}, {-0.75+\j}) -- ({0.5*\x+2*\i}, {-1+\j}); \draw[BrickRed,thick] ({-0.5*\x+2*\i}, {0.75+\j}) -- ({-0.5*\x+2*\i}, {1+\j}); \draw[BrickRed,thick] ({0.5*\x+2*\i}, {0.75+\j}) -- ({0.5*\x+2*\i}, {1+\j});
}
\draw[white,fill=white] ({(-0.75-3.02)*\x}, -1.25) rectangle ({(-0.75-2.25)*\x}, 1.25) {}; \draw[white,fill=white] ({(0.75+3.02)*\x}, -1.25) rectangle ({(0.75+2.25)*\x}, 1.25) {};
\end{tikzpicture} ,
\end{equation}
that uses a PEPS tensor of bond dimension equal to two, and a physical space gathering six atoms forming a bowtie. This tensor only has eight non-zero elements, all equal to one, that can be depicted as 
\begin{equation} \label{eq_AllNonzeroCoefficients} \begin{split}
& 
%%%%%%%% Coefficient 1
%
\begin{tikzpicture}[scale=0.4,baseline=-1,every node/.style={scale=0.8}]
\def\x{sqrt(3)}	
\draw[rounded corners] ({-0.75*\x}, -0.75) rectangle ({0.75*\x}, 0.75) {};
\draw[thick] ({0.5*\x},-0.5) -- ({-0.5*\x},0.5) -- ({-0.5*\x},-0.5) -- ({0.5*\x},0.5) -- ({0.5*\x},-0.5); 
\draw[PineGreen,fill=PineGreen] ({0.25*\x},0.25) circle (0.15); 
\draw[BrickRed,thick] ({-0.5*\x}, 0.5) -- ({-0.5*\x}, 1) node[above] {0}; \draw[BrickRed,thick] ({0.5*\x}, 0.5) -- ({0.5*\x}, 1) node[above] {$-$};
\draw[BrickRed,thick] ({-0.5*\x}, -0.5) -- ({-0.5*\x}, -1) node[below] {$-$}; \draw[BrickRed,thick] ({0.5*\x}, -0.5) -- ({0.5*\x}, -1) node[below] {1}; \end{tikzpicture} , \quad
%
%%%%%%%% Coefficient 2
%
\begin{tikzpicture}[scale=0.4,baseline=-1,every node/.style={scale=0.8}]
\def\x{sqrt(3)}	
\draw[rounded corners] ({-0.75*\x}, -0.75) rectangle ({0.75*\x}, 0.75) {};
\draw[thick] ({0.5*\x},-0.5) -- ({-0.5*\x},0.5) -- ({-0.5*\x},-0.5) -- ({0.5*\x},0.5) -- ({0.5*\x},-0.5); 
\draw[PineGreen,fill=PineGreen] ({0.25*\x},-0.25) circle (0.15); 
\draw[BrickRed,thick] ({-0.5*\x}, 0.5) -- ({-0.5*\x}, 1) node[above] {0}; \draw[BrickRed,thick] ({0.5*\x}, 0.5) -- ({0.5*\x}, 1) node[above] {+};
\draw[BrickRed,thick] ({-0.5*\x}, -0.5) -- ({-0.5*\x}, -1) node[below] {$-$}; \draw[BrickRed,thick] ({0.5*\x}, -0.5) -- ({0.5*\x}, -1) node[below] {0};
\end{tikzpicture} , \quad
%
%%%%%%%% Coefficient 3
%
\begin{tikzpicture}[scale=0.4,baseline=-1,every node/.style={scale=0.8}]
\def\x{sqrt(3)}	
\draw[rounded corners] ({-0.75*\x}, -0.75) rectangle ({0.75*\x}, 0.75) {};
\draw[thick] ({0.5*\x},-0.5) -- ({-0.5*\x},0.5) -- ({-0.5*\x},-0.5) -- ({0.5*\x},0.5) -- ({0.5*\x},-0.5); 
\draw[PineGreen,fill=PineGreen] ({-0.25*\x},0.25) circle (0.15); 
\draw[BrickRed,thick] ({-0.5*\x}, 0.5) -- ({-0.5*\x}, 1) node[above] {1}; \draw[BrickRed,thick] ({0.5*\x}, 0.5) -- ({0.5*\x}, 1) node[above] {+};
\draw[BrickRed,thick] ({-0.5*\x}, -0.5) -- ({-0.5*\x}, -1) node[below] {$-$}; \draw[BrickRed,thick] ({0.5*\x}, -0.5) -- ({0.5*\x}, -1) node[below] {1};
\end{tikzpicture} , \quad
%
%%%%%%%% Coefficient 4
%
\begin{tikzpicture}[scale=0.4,baseline=-1,every node/.style={scale=0.8}]
\def\x{sqrt(3)}	
\draw[rounded corners] ({-0.75*\x}, -0.75) rectangle ({0.75*\x}, 0.75) {};
\draw[thick] ({0.5*\x},-0.5) -- ({-0.5*\x},0.5) -- ({-0.5*\x},-0.5) -- ({0.5*\x},0.5) -- ({0.5*\x},-0.5); 
\draw[PineGreen,fill=PineGreen] ({-0.25*\x},-0.25) circle (0.15); 
\draw[BrickRed,thick] ({-0.5*\x}, 0.5) -- ({-0.5*\x}, 1) node[above] {0}; \draw[BrickRed,thick] ({0.5*\x}, 0.5) -- ({0.5*\x}, 1) node[above] {+};
\draw[BrickRed,thick] ({-0.5*\x}, -0.5) -- ({-0.5*\x}, -1) node[below] {+}; \draw[BrickRed,thick] ({0.5*\x}, -0.5) -- ({0.5*\x}, -1) node[below] {1};
\end{tikzpicture} , \\
& 
%%%%%%%% Coefficient 5
%
\begin{tikzpicture}[scale=0.4,baseline=-1,every node/.style={scale=0.8}]
\def\x{sqrt(3)}	
\draw[rounded corners] ({-0.75*\x}, -0.75) rectangle ({0.75*\x}, 0.75) {};
\draw[thick] ({0.5*\x},-0.5) -- ({-0.5*\x},0.5) -- ({-0.5*\x},-0.5) -- ({0.5*\x},0.5) -- ({0.5*\x},-0.5); 
\draw[PineGreen,fill=PineGreen] ({0.25*\x},0.25) circle (0.15); 
\draw[PineGreen,fill=PineGreen] ({-0.5*\x},0) circle (0.15); 
\draw[BrickRed,thick] ({-0.5*\x}, 0.5) -- ({-0.5*\x}, 1) node[above] {1}; \draw[BrickRed,thick] ({0.5*\x}, 0.5) -- ({0.5*\x}, 1) node[above] {$-$};
\draw[BrickRed,thick] ({-0.5*\x}, -0.5) -- ({-0.5*\x}, -1) node[below] {+}; \draw[BrickRed,thick] ({0.5*\x}, -0.5) -- ({0.5*\x}, -1) node[below] {1}; \end{tikzpicture} , \quad
%
%%%%%%%% Coefficient 6
%
\begin{tikzpicture}[scale=0.4,baseline=-1,every node/.style={scale=0.8}]
\def\x{sqrt(3)}	
\draw[rounded corners] ({-0.75*\x}, -0.75) rectangle ({0.75*\x}, 0.75) {};
\draw[thick] ({0.5*\x},-0.5) -- ({-0.5*\x},0.5) -- ({-0.5*\x},-0.5) -- ({0.5*\x},0.5) -- ({0.5*\x},-0.5); 
\draw[PineGreen,fill=PineGreen] ({0.25*\x},-0.25) circle (0.15); 
\draw[PineGreen,fill=PineGreen] ({-0.5*\x},0) circle (0.15); 
\draw[BrickRed,thick] ({-0.5*\x}, 0.5) -- ({-0.5*\x}, 1) node[above] {1}; \draw[BrickRed,thick] ({0.5*\x}, 0.5) -- ({0.5*\x}, 1) node[above] {+};
\draw[BrickRed,thick] ({-0.5*\x}, -0.5) -- ({-0.5*\x}, -1) node[below] {+}; \draw[BrickRed,thick] ({0.5*\x}, -0.5) -- ({0.5*\x}, -1) node[below] {0};
\end{tikzpicture} , \quad
%
%%%%%%%% Coefficient 7
%
\begin{tikzpicture}[scale=0.4,baseline=-1,every node/.style={scale=0.8}]
\def\x{sqrt(3)}	
\draw[rounded corners] ({-0.75*\x}, -0.75) rectangle ({0.75*\x}, 0.75) {};
\draw[thick] ({0.5*\x},-0.5) -- ({-0.5*\x},0.5) -- ({-0.5*\x},-0.5) -- ({0.5*\x},0.5) -- ({0.5*\x},-0.5); 
\draw[PineGreen,fill=PineGreen] ({-0.25*\x},0.25) circle (0.15); 
\draw[PineGreen,fill=PineGreen] ({0.5*\x},0) circle (0.15); 
\draw[BrickRed,thick] ({-0.5*\x}, 0.5) -- ({-0.5*\x}, 1) node[above] {1}; \draw[BrickRed,thick] ({0.5*\x}, 0.5) -- ({0.5*\x}, 1) node[above] {$-$};
\draw[BrickRed,thick] ({-0.5*\x}, -0.5) -- ({-0.5*\x}, -1) node[below] {$-$}; \draw[BrickRed,thick] ({0.5*\x}, -0.5) -- ({0.5*\x}, -1) node[below] {0};
\end{tikzpicture} , \quad
%
%%%%%%%% Coefficient 8
%
\begin{tikzpicture}[scale=0.4,baseline=-1,every node/.style={scale=0.8}]
\def\x{sqrt(3)}	
\draw[rounded corners] ({-0.75*\x}, -0.75) rectangle ({0.75*\x}, 0.75) {};
\draw[thick] ({0.5*\x},-0.5) -- ({-0.5*\x},0.5) -- ({-0.5*\x},-0.5) -- ({0.5*\x},0.5) -- ({0.5*\x},-0.5); 
\draw[PineGreen,fill=PineGreen] ({-0.25*\x},-0.25) circle (0.15); 
\draw[PineGreen,fill=PineGreen] ({0.5*\x},0) circle (0.15); 
\draw[BrickRed,thick] ({-0.5*\x}, 0.5) -- ({-0.5*\x}, 1) node[above] {0}; \draw[BrickRed,thick] ({0.5*\x}, 0.5) -- ({0.5*\x}, 1) node[above] {$-$};
\draw[BrickRed,thick] ({-0.5*\x}, -0.5) -- ({-0.5*\x}, -1) node[below] {+}; \draw[BrickRed,thick] ({0.5*\x}, -0.5) -- ({0.5*\x}, -1) node[below] {0};
\end{tikzpicture} , 
\end{split} \end{equation}
with filled circle representing atoms in $\ket{e}$, and where we have used the notation $\ket{\pm} = (\ket{0} \pm \ket{1})/\sqrt{2}$ in the auxiliary space. The two upper auxiliary qubits are equal to ones or pluses if the vertex closest to them is covered by a dimer, and to zero or minus otherwise. The opposite is true for the two lower auxiliary qubits, such that, upon contraction, only configurations with all vertices covered by one and exactly one dimer acquire a non-zero weight. Since all coefficients in Eq.~\ref{eq_AllNonzeroCoefficients} are equal to one, this weight also is, and Eq.~\ref{eq_TotalStateOnLattice} indeed represents the dimer state. Note that, because we kept the physical degrees of freedom of the tensors on the atomic sites rather than on the vertices of the Kagome lattice, our construction differs from, but is equivalent to, earlier PEPS descriptions of dimer states~\cite{verstraete2006criticality,schuch2012resonating}.

Logical qubits are encoded into the auxiliary space of the tensors, and form computational wires (red vertical lines in Eq.~\ref{eq_TotalStateOnLattice}). In the remainder of the discussion, we provide an explicit way to devise a universal set of gates on these logical qubits that solely rely on the pulses (I-III) given above. The quantum computation is run from top to bottom by application of these pulses followed by measurements of the physical leg of the PEPS tensor forming $\ket{\Phi}$. Note that because the logical qubits are encoded on the state of six atoms (see Sec.~\ref{eq_AllNonzeroCoefficients}), the measurements (I-III) also involve the measurement of multiple atoms in order to change the state of one logical qubit.

\subsection{Wire decoupling} \label{ssec_wiredecoupling}

First, logical qubits should be able to pass through PEPS tensors without being altered. In other word, we should be able to propagate quantum information until the next gate of our algorithm is required. To achieve this, we shall find a measurement sequence that decouples the two computational wire incoming in each of the PEPS tensor. The following procedure accomplishes this purpose: the right and leftmost atoms of the bowtie are first isolated and measured in the $\ket{g} \pm \ket{e}$ basis (\textit{i.e.} fluorescence after a $\pi/2$-pulse); then, a $\varphi = -\pi/2$ phase shift is imprinted on the two bottom atoms of the bowtie's central cross (III); finally, the atoms of the central cross are gathered two-by-two vertically, each within its own blockade radius, and pulse (I) is performed.

Our above analysis shows that this sequence projects the PEPS' physical state onto one of the states
\begin{equation} \label{eq_PhysicalStatesDecouplingWires} \begin{cases}
\left( \begin{tikzpicture}[scale=0.4,baseline=-2.5,every node/.style={scale=0.8}]
\draw[thick] (0,-0.5) -- (0,0.5); 
\end{tikzpicture} +x_L \begin{tikzpicture}[scale=0.4,baseline=-2.5,every node/.style={scale=0.8}]
\draw[thick] (0,-0.5) -- (0,0.5); 
\draw[PineGreen,fill=PineGreen] (0.,0.) circle (0.15); 
\end{tikzpicture}  \right) \times 
\left( \begin{tikzpicture}[scale=0.4,baseline=-2.5,every node/.style={scale=0.8}]
\def\x{sqrt(3)}	
\draw[thick] (0.,0.) -- ({-0.5*\x},0.5); \draw[thick] (0.,0.) -- ({-0.5*\x},-0.5); 
\draw[PineGreen,fill=PineGreen] ({-0.25*\x},0.25) circle (0.15); 
\end{tikzpicture} +p_L \begin{tikzpicture}[scale=0.4,baseline=-2.5,every node/.style={scale=0.8}]
\def\x{sqrt(3)}	
\draw[thick] (0.,0.) -- ({-0.5*\x},0.5); \draw[thick] (0.,0.) -- ({-0.5*\x},-0.5); 
\draw[PineGreen,fill=PineGreen] ({-0.25*\x},-0.25) circle (0.15); 
\end{tikzpicture}  \right) \times 
\begin{tikzpicture}[scale=0.4,baseline=-2.5,every node/.style={scale=0.8}]
\def\x{sqrt(3)}	
\draw[thick] ({0.5*\x},-0.5) -- (0.,0.) -- ({0.5*\x},0.5); 
\end{tikzpicture}
\times
\left( \begin{tikzpicture}[scale=0.4,baseline=-2.5,every node/.style={scale=0.8}]
\draw[thick] (0,-0.5) -- (0,0.5); 
\end{tikzpicture} +x_R \begin{tikzpicture}[scale=0.4,baseline=-2.5,every node/.style={scale=0.8}]
\draw[thick] (0,-0.5) -- (0,0.5); 
\draw[PineGreen,fill=PineGreen] (0.,0.) circle (0.15); 
\end{tikzpicture}  \right)  & \text{if: } p_R=0 \\
\left( \begin{tikzpicture}[scale=0.4,baseline=-2.5,every node/.style={scale=0.8}]
\draw[thick] (0,-0.5) -- (0,0.5); 
\end{tikzpicture} +x_L \begin{tikzpicture}[scale=0.4,baseline=-2.5,every node/.style={scale=0.8}]
\draw[thick] (0,-0.5) -- (0,0.5); 
\draw[PineGreen,fill=PineGreen] (0.,0.) circle (0.15); 
\end{tikzpicture}  \right) \times 
\begin{tikzpicture}[scale=0.4,baseline=-2.5,every node/.style={scale=0.8}]
\def\x{sqrt(3)}	
\draw[thick] ({-0.5*\x},-0.5) -- (0.,0.) -- ({-0.5*\x},0.5); 
\end{tikzpicture} \times 
\left( \begin{tikzpicture}[scale=0.4,baseline=-2.5,every node/.style={scale=0.8}]
\def\x{sqrt(3)}	
\draw[thick] ({0.5*\x},-0.5) -- (0.,0.); \draw[thick] ({0.5*\x},0.5) -- (0.,0.); 
\draw[PineGreen,fill=PineGreen] ({0.25*\x},0.25) circle (0.15); 
\end{tikzpicture} + p_R \begin{tikzpicture}[scale=0.4,baseline=-2.5,every node/.style={scale=0.8}]
\def\x{sqrt(3)}	
\draw[thick] ({0.5*\x},-0.5) -- (0.,0.); \draw[thick] ({0.5*\x},0.5) -- (0.,0.); 
\draw[PineGreen,fill=PineGreen] ({0.25*\x},-0.25) circle (0.15); 
\end{tikzpicture}  \right) \times
\left( \begin{tikzpicture}[scale=0.4,baseline=-2.5,every node/.style={scale=0.8}]
\draw[thick] (0,-0.5) -- (0,0.5); 
\end{tikzpicture} +x_R \begin{tikzpicture}[scale=0.4,baseline=-2.5,every node/.style={scale=0.8}]
\draw[thick] (0,-0.5) -- (0,0.5); 
\draw[PineGreen,fill=PineGreen] (0.,0.) circle (0.15); 
\end{tikzpicture}  \right)  & \text{if: } p_L=0 
\end{cases} , \end{equation}
where $x_L, x_R= \pm$ are the measurement results from the right/left-most atoms, and $p_L, p_R$ are those of the left and right two-atoms clusters, as defined after Eq.~\ref{eq_EffectiveMeasurement}. Using Eq.~\ref{eq_AllNonzeroCoefficients} (see App.~\ref{app_gate}), we can then compute the action that this projection, written as $D_0(x_L,p_L,p_R,x_R)$, has on the incoming logical qubit
\begin{equation} \label{eq_GatePP}
\begin{tikzpicture}[scale=0.8,baseline=-2.5]
\def\x{sqrt(3)}	; \draw[rounded corners] ({-1.1*\x}, -0.5) rectangle ({1.1*\x}, 0.5) {}; \node at (0,0) {$D_0(x_L,p_L,p_R,x_R)$} ;
\draw[BrickRed,thick] ({-0.5*\x}, -0.5) -- ({-0.5*\x}, -1); \draw[BrickRed,thick] ({0.5*\x}, -0.5) -- ({0.5*\x}, -1); \draw[BrickRed,thick] ({-0.5*\x}, 0.5) -- ({-0.5*\x}, 1); \draw[BrickRed,thick] ({0.5*\x}, 0.5) -- ({0.5*\x}, 1);
\end{tikzpicture}  \equiv  
\begin{tikzpicture}[scale=0.8,baseline=-2.5]
\draw[BrickRed,thick] (-1.3, 1) -- (-1.3, -1); \draw[fill=white] (-0.8, 0.9) rectangle (-1.8, 0.4) node[pos=.5] {$H$}; \draw[fill=white] (-1.8, -0.25) rectangle (-0.8, 0.25) node[pos=.5] {$Z^{|p_R|}$}; \draw[fill=white] (-2.5, -0.9) rectangle (-0.1, -0.4) node[pos=.5] {$X^{p_L \tilde{p}_L + p_R \tilde{x}_L}$};
\draw[BrickRed,thick] (1.3, 1) -- (1.3, -1); \draw[fill=white] (1.8, 0.9) rectangle (0.8, 0.4) node[pos=.5] {$H$}; \draw[fill=white] (2.5, 0.25) rectangle (0.1, -0.25) node[pos=.5] {$Z^{p_R \tilde{p}_R + p_L \tilde{x}_R}$}; \draw[fill=white] (1.8, -0.4) rectangle (0.8, -0.9) node[pos=.5] {$X^{|p_L|}$};
\end{tikzpicture} \, ,
\end{equation}
where the tilde variable are defined as $\tilde{q} = (1-q)/2$. The two last lines are irrelevant Pauli errors that are inevitable within MBQC, and can be corrected at the end of the computation. We observe that the computational wires are indeed decoupled by $D_0$, up to Hadamard gates that will cancel out with those arising in adjacent PEPS tensors (see below).

\subsection{Single qubit operations} \label{ssec_singlequbitgate}

A small modification of the previous scheme allows to implement single qubit rotations along each of the incoming computation wires. More precisely, if the leftmost atom of the bowtie is measured in $\ket{g} \pm e^{i\varphi}\ket{e}$ using (III), the gate implemented changes as follows
\begin{equation} \label{eq_GatePtheta}
\begin{tikzpicture}[scale=0.8,baseline=-2.5]
\def\x{sqrt(3)}	; \draw[rounded corners] ({-0.5*\x}, -0.5) rectangle ({0.5*\x}, 0.5) {}; \node at (0,0) {$D_\varphi$} ;
\draw[BrickRed,thick] ({-0.3*\x}, -0.5) -- ({-0.3*\x}, -1); \draw[BrickRed,thick] ({0.3*\x}, -0.5) -- ({0.3*\x}, -1); \draw[BrickRed,thick] ({-0.3*\x}, 0.5) -- ({-0.3*\x}, 1); \draw[BrickRed,thick] ({0.3*\x}, 0.5) -- ({0.3*\x}, 1);
\end{tikzpicture}  \equiv  
\begin{tikzpicture}[scale=0.8,baseline=-2.5]
\draw[BrickRed,thick] (-1.2, 1) -- (-1.2, -1); \draw[fill=white] (-0.8, 0.9) rectangle (-1.6, 0.4) node[pos=.5] {$H$}; \draw[fill=white] (-2.3, -0.9) rectangle (-0.1, -0.4) node[pos=.5] {$X^{p_L+x_L p_R}$}; \draw[fill=white] (-1.8, 0.3) rectangle (-0.6, -0.3) node[pos=.5] {$G_\varphi^{|p_R|}$};
\draw[BrickRed,thick] (1.2, 1) -- (1.2, -1); \draw[fill=white] (1.6, 0.9) rectangle (0.8, 0.4) node[pos=.5] {$H$}; \draw[fill=white] (2.3, 0.25) rectangle (0.1, -0.25) node[pos=.5] {$Z^{p_R+x_R p_L}$}; \draw[fill=white] (1.6, -0.4) rectangle (0.8, -0.9) node[pos=.5] {$X^{p_L}$};
\end{tikzpicture} \, , \quad G_\varphi = Z R_x(\varphi) ,
\end{equation}
where $R_x$ denotes a rotation around the $x$-axis on the logical qubit's Bloch sphere. The conditional rotation $G_\varphi$ is similarly performed on the right computational wire when the rightmost atom of the bowtie is phase-shifted instead of the leftmost one.

While this rotation is only applied when $p_R \neq 0$, we can try to implement it on successive PEPS tensor along the computational wire until a non-zero $p_R$ heralds successful rotation of the qubit. We now show that the probability to measure $p_R = 0$ on $n$ successive tensors is $K / 2^{n-1}$, with $K$ constant, ensuring the quasi-certain application of the rotation $G_\varphi$ with polynomial circuit depth using the above iterative method.

To prove this statement, consider the following situation 
\begin{equation}
\begin{tikzpicture}[scale=0.5,baseline=(current bounding box.center)]
\def\x{sqrt(3)}; \draw[BrickRed,thick] ({-\x},1.) -- ({-\x},-2); \draw[BrickRed,thick] ({\x},1) -- ({\x},-2); \draw[BrickRed,thick] (0,1) -- (0,-2);
\draw[thick] ({\x},-0.5) -- (0,0.5) -- (0,-0.5) -- ({\x},0.5); \draw[thick] ({-\x},-0.5) -- (0,-1.5) -- (0,-0.5) -- ({-\x},-1.5);
\draw[PineGreen] ({0.25*\x},0.25) circle (0.15) node[above] {1};
\draw[PineGreen] ({0.25*\x},-0.25) circle (0.15) node[below right] {2};
\draw[PineGreen] (0.,-1.) circle (0.15) node[below right] {3};
\draw[PineGreen] ({-0.25*\x},-0.75) circle (0.15) node[above] {4};
\draw[PineGreen] ({-0.25*\x},-1.25) circle (0.15) node[below] {5};
\end{tikzpicture} \; ,
\end{equation}
and assume that we want to apply a rotation $G_\varphi$ on the central logical wire but have measured $p_R=0$ on the upper bowtie. We thus apply a phase shift on the rightmost atom of the lower bowtie, and would like to know the probability of success ($p_L \neq 0$). The measurement $p_R = 0$ has restricted $\ket{\Phi}$ to an equal superposition of dimer covering in which either atom 1 or atom 2 is excited. Suppose it is atom 1, then either atom 3 or 4 should also be excited to satisfy the dimer constraint, which respectively yields $p_L \neq 0$ and $p_L = 0$. We can see that the number of dimer coverings with excitations on 1-3 and 1-4 is identical. Indeed, there exists a bijection $B$ between these dimer covering subsets~\cite{verresen2021prediction}
\begin{equation}
\begin{tikzpicture}[scale=0.5,baseline=(current bounding box.center)]
\def\x{sqrt(3)}; \draw[BrickRed,thick] ({-\x},1.) -- ({-\x},-2); \draw[BrickRed,thick] ({\x},1) -- ({\x},-2); \draw[BrickRed,thick] (0,1) -- (0,-2);
\draw[thick] ({\x},-0.5) -- (0,0.5) -- (0,-0.5) -- ({\x},0.5); \draw[thick] ({-\x},-0.5) -- (0,-1.5) -- (0,-0.5) -- ({-\x},-1.5);
\draw[PineGreen] ({0.25*\x},0.25) circle (0.15) node[above] {1};
\draw[PineGreen] (0.,-1.) circle (0.15) node[right] {3};
\draw[PineGreen] ({-0.25*\x},-0.75) circle (0.15) node[above] {4};
\draw[thick]  ({-\x},-1.5) -- ({-\x},-2.5) -- ({-0.5*\x},-3) -- (0,-2.5) -- (0,-1.5); \draw[thick]  ({-\x},-1.5) -- ({-1.5*\x},-2) -- ({-\x},-2.5) -- ({-\x},-3.5) -- ({-0.5*\x},-3) -- (0,-3.5) -- (0,-2.5) -- ({0.5*\x},-2) -- (0,-1.5);
\draw[RoyalBlue,thick,decorate,decoration={snake,amplitude=.4mm,segment length=0.59mm,post length=0mm}] ({-\x},-1.5) -- ({-\x},-2.5) -- ({-0.5*\x},-3) -- (0,-2.5) -- (0,-1.5) -- ({-0.5*\x},-1) -- ({-\x},-1.5); \node[RoyalBlue] at ({-0.5*\x},-2) {$B$};
\end{tikzpicture} \; , 
\quad \begin{tikzpicture}[scale=0.5,baseline=-2.5]
\def\x{sqrt(3)}; \draw[thick] ({0.5*\x},0.) -- (0,0.5) -- (0,-0.5) -- ({0.5*\x},0.);
\draw[RoyalBlue,thick,decorate,decoration={snake,amplitude=.4mm,segment length=0.59mm,post length=0mm}] (0,0.5) -- (0,-0.5); 
\end{tikzpicture} : \left\{
\begin{array}{rcr}
\begin{tikzpicture}[scale=0.5,baseline=-2.5]
\def\x{sqrt(3)}; \draw[thick] ({0.5*\x},0.) -- (0,0.5) -- (0,-0.5) -- ({0.5*\x},0.);
\end{tikzpicture} & \leftrightarrows & \begin{tikzpicture}[scale=0.5,baseline=-2.5]
\def\x{sqrt(3)}; \draw[thick] ({0.5*\x},0.) -- (0,0.5) -- (0,-0.5) -- ({0.5*\x},0.); \draw[PineGreen,fill=PineGreen] (0,0) circle (0.15);
\end{tikzpicture} \\
\begin{tikzpicture}[scale=0.5,baseline=-2.5]
\def\x{sqrt(3)}; \draw[thick] ({0.5*\x},0.) -- (0,0.5) -- (0,-0.5) -- ({0.5*\x},0.); \draw[PineGreen,fill=PineGreen] ({0.25*\x},-0.25) circle (0.15);
\end{tikzpicture} & \leftrightarrows & \begin{tikzpicture}[scale=0.5,baseline=-2.5]
\def\x{sqrt(3)}; \draw[thick] ({0.5*\x},0.) -- (0,0.5) -- (0,-0.5) -- ({0.5*\x},0.); \draw[PineGreen,fill=PineGreen] ({0.25*\x},0.25) circle (0.15);
\end{tikzpicture} 
\end{array}
\right. ,
\end{equation}
where we have defined $B$ by its action on each of the triangles. A similar analysis holds when atom 2 is in $\ket{e}$. As a result, there are as many dimer coverings leading to $p_L = 0$ and $p_L \neq 0$ once $p_R = 0$ has been measured. Since $\ket{\Phi}$ is an equal superposition of those, the probability of both outcomes is identical, equal to one half. If we write $K$ the probability to measure $p_R = 0$ during the first trial, the probability of failing to apply $G_\varphi$ on $n$ consecutive PEPS tensors is thus $K/2^{n-1}$, as claimed above.

Finally, notice that $\varphi$ can be tuned to realize, among others, the three gates $G_0 = Z$, $G_{\pi/2} = Z\sqrt{X}$ and $G_\pi = ZX$, which form a universal gate set for single qubit operations since $\sqrt{X}$ is a non-Clifford gate~\cite{nebe2001invariants}. We have therefore demonstrated that it is possible to implement arbitrary single qubit operations along the computational wires of our system with a polynomial circuit depth.

\subsection{Universal quantum computation}  \label{ssec_twoqubitgate}

To reach universality, the current gate set should be expanded with a two-qubit gate that is not unitarily equivalent to a SWAP. We propose the following sequence, named $Q$: Three of the bowtie's atoms are dephased by $\pi$ (III), for instance, the rightmost atoms and those next to the leftmost one; then pulse (II) is applied to the left and right triangles. This projects the bowtie's physical space onto 
\begin{equation} \label{eq_twoqubitgateprojection} \begin{split}
& \left(
\begin{tikzpicture}[scale=0.3,baseline=-2.7]
\def\x{sqrt(3)}; \draw[thick] ({0.5*\x},0) -- (0,0.5) -- (0,-0.5) -- ({0.5*\x},0); 
\end{tikzpicture} - p_\uparrow^L p_\downarrow^L \begin{tikzpicture}[scale=0.3,baseline=-2.7]
\def\x{sqrt(3)}; \draw[thick] ({0.5*\x},0) -- (0,0.5) -- (0,-0.5) -- ({0.5*\x},0); \draw[PineGreen,fill=PineGreen] (0,0) circle (0.2); 
\end{tikzpicture} + x^L p_\uparrow^L \begin{tikzpicture}[scale=0.3,baseline=-2.7]
\def\x{sqrt(3)}; \draw[thick] ({0.5*\x},0) -- (0,0.5) -- (0,-0.5) -- ({0.5*\x},0); \draw[PineGreen,fill=PineGreen] ({0.25*\x},-0.25) circle (0.2); 
\end{tikzpicture} + x^L p_\downarrow^L \begin{tikzpicture}[scale=0.3,baseline=-2.7]
\def\x{sqrt(3)}; \draw[thick] ({0.5*\x},0) -- (0,0.5) -- (0,-0.5) -- ({0.5*\x},0); \draw[PineGreen,fill=PineGreen] ({0.25*\x},0.25) circle (0.2); 
\end{tikzpicture} \right) \\
& \qquad \times \left(
\begin{tikzpicture}[scale=0.3,baseline=-2.7]
\def\x{-sqrt(3)}; \draw[thick] ({0.5*\x},0) -- (0,0.5) -- (0,-0.5) -- ({0.5*\x},0); 
\end{tikzpicture} + p_\uparrow^R p_\downarrow^R \begin{tikzpicture}[scale=0.3,baseline=-2.7]
\def\x{-sqrt(3)}; \draw[thick] ({0.5*\x},0) -- (0,0.5) -- (0,-0.5) -- ({0.5*\x},0); \draw[PineGreen,fill=PineGreen] (0,0) circle (0.2); 
\end{tikzpicture} - x^R p_\uparrow^R \begin{tikzpicture}[scale=0.3,baseline=-2.7]
\def\x{-sqrt(3)}; \draw[thick] ({0.5*\x},0) -- (0,0.5) -- (0,-0.5) -- ({0.5*\x},0); \draw[PineGreen,fill=PineGreen] ({0.25*\x},-0.25) circle (0.2); 
\end{tikzpicture} - x^R p_\downarrow^R \begin{tikzpicture}[scale=0.3,baseline=-2.7]
\def\x{-sqrt(3)}; \draw[thick] ({0.5*\x},0) -- (0,0.5) -- (0,-0.5) -- ({0.5*\x},0); \draw[PineGreen,fill=PineGreen] ({0.25*\x},0.25) circle (0.2); 
\end{tikzpicture} \right) ,
\end{split} \end{equation}
where the $x$, $p_\uparrow$ and $p_\downarrow$'s have been defined above. Using Eq.~\ref{eq_AllNonzeroCoefficients}, one can check that the PEPS tensor obtained after contraction of the physical index acts on the two incoming logical qubits as
\begin{equation}
\begin{tikzpicture}[scale=0.8,baseline=-2.5]
\def\x{sqrt(3)}	; \draw[rounded corners] ({-0.5*\x}, -0.5) rectangle ({0.5*\x}, 0.5) {}; \node at (0,0) {$Q$} ;
\draw[BrickRed,thick] ({-0.3*\x}, -0.5) -- ({-0.3*\x}, -1); \draw[BrickRed,thick] ({0.3*\x}, -0.5) -- ({0.3*\x}, -1); \draw[BrickRed,thick] ({-0.3*\x}, 0.5) -- ({-0.3*\x}, 1); \draw[BrickRed,thick] ({0.3*\x}, 0.5) -- ({0.3*\x}, 1);
\end{tikzpicture}  \equiv  
\begin{tikzpicture}[scale=0.8,baseline=-2.5]
\def\x{sqrt(3)}; \draw[BrickRed,thick] ({-0.3*\x}, 1) -- ({-0.3*\x}, -1); \draw[BrickRed,thick] ({0.3*\x}, 1) -- ({0.3*\x}, -1); 
\draw[fill=white] ({0.3*\x-0.4}, 0.9) rectangle ({0.3*\x+0.4}, 0.4) node[pos=.5] {$H$}; \draw[fill=white] ({-0.3*\x-0.4}, 0.9) rectangle ({-0.3*\x+0.4}, 0.4) node[pos=.5] {$H$};
\draw ({-0.3*\x}, 0.15) -- ({0.3*\x+0.15}, 0.15); \draw[fill=black] ({-0.3*\x},0.15) circle (0.15); \draw[black] ({0.3*\x},0.15) circle (0.15); 
\draw[fill=white] ({-0.3*\x-0.4}, -0.05) rectangle ({-0.3*\x+0.4}, -0.55) node[pos=.5] {$H$}; \draw[fill=white] ({0.3*\x-0.4}, -0.05) rectangle ({0.3*\x+0.4}, -0.55) node[pos=.5] {$X$};
\draw ({-0.3*\x}, -0.75) -- ({0.3*\x+0.15}, -0.75); \draw[fill=black] ({-0.3*\x},-0.75) circle (0.15); \draw[black] ({0.3*\x},-0.75) circle (0.15); 
\end{tikzpicture} , 
\end{equation}
up to irrelevant Pauli errors (see App.~\ref{app_gate}). Apart from the initial Hadamard gates common to all of our gate set, $Q$ can be understood as an entangling Bell gate that sends, for instance, $\ket{00} \to \ket{01} + \ket{10}$. Such a gate creates entanglement and therefore provides universality when appended to our previous single qubit operations.

\section{State preparation imperfection} \label{sec_imperfections}

For the moment, we have assumed that the Kagome dimer state $\ket{\Phi}$ perfectly describes the state realized in the Rydberg array of Ref.~\cite{semeghini2021probing}. While, as stated above, this approximation is justified by extensive numerical works~\cite{verresen2021efficiently,verresen2021prediction,giudici2022dynamical}, we may nevertheless wonder how sensitive is the universal MBQC scheme presented in the last section to small imperfections in the experimental realization of the Kagome dimer state.

\subsection{Modeling deviations from the dimer state}

Since the goal of this paper is neither to provide a detailed account of all noise sources of Ref.~\cite{semeghini2021probing} nor to optimize dynamical preparation protocols for $\ket{\Phi}$, we will model typical deviations away from the Kagome dimer state and estimate the gate fidelity resulting from these deviations. Similarly, we will not consider errors related to imprecise measurements, as these errors can be corrected to a large degree with pulse engineering. Finally, we will focus on a specific form of deviations from the dimer state, in which each PEPS tensor forming $\ket{\Phi}$ can be modified 
\begin{equation}
\begin{tikzpicture}[scale=0.4,baseline=-1]
\def\x{sqrt(3)}	
\draw[rounded corners] ({(-0.75)*\x}, {-0.75}) rectangle ({(0.75)*\x}, {0.75}) {};
\node at (0,0) {$P$};
\draw[BrickRed,thick] ({-0.5*\x}, {-0.75}) -- ({-0.5*\x}, {-1}); \draw[BrickRed,thick] ({0.5*\x}, {-0.75}) -- ({0.5*\x}, {-1}); \draw[BrickRed,thick] ({-0.5*\x}, {0.75}) -- ({-0.5*\x}, {1}); \draw[BrickRed,thick] ({0.5*\x}, {0.75}) -- ({0.5*\x}, {1});
\end{tikzpicture}
=
\begin{tikzpicture}[scale=0.4,baseline=-1]
\def\x{sqrt(3)}	
\draw[rounded corners] ({(-0.75)*\x}, {-0.75}) rectangle ({(0.75)*\x}, {0.75}) {};
\draw[thick] ({0.5*\x},{-0.5}) -- ({-0.5*\x},{0.5}) -- ({-0.5*\x},{-0.5}) -- ({0.5*\x},{0.5}) -- ({0.5*\x},{-0.5}); 
\draw[BrickRed,thick] ({-0.5*\x}, {-0.75}) -- ({-0.5*\x}, {-1}); \draw[BrickRed,thick] ({0.5*\x}, {-0.75}) -- ({0.5*\x}, {-1}); \draw[BrickRed,thick] ({-0.5*\x}, {0.75}) -- ({-0.5*\x}, {1}); \draw[BrickRed,thick] ({0.5*\x}, {0.75}) -- ({0.5*\x}, {1});
\end{tikzpicture}
+ \eta \, 
\begin{tikzpicture}[scale=0.4,baseline=-1]
\def\x{sqrt(3)}	
\draw[rounded corners] ({(-0.75)*\x}, {-0.75}) rectangle ({(0.75)*\x}, {0.75}) {};
\node at (0,0) {$\delta P$};
\draw[BrickRed,thick] ({-0.5*\x}, {-0.75}) -- ({-0.5*\x}, {-1}); \draw[BrickRed,thick] ({0.5*\x}, {-0.75}) -- ({0.5*\x}, {-1}); \draw[BrickRed,thick] ({-0.5*\x}, {0.75}) -- ({-0.5*\x}, {1}); \draw[BrickRed,thick] ({0.5*\x}, {0.75}) -- ({0.5*\x}, {1});
\end{tikzpicture} ,
\end{equation}
with $\eta \ll 1$ and $\delta P$ an arbitrary tensor. This form allows for efficient computations of gate fidelities, but at the same time carries a great representative power due to the large dimension of $\delta P$ -- which has four two dimensional external legs and a $2^6$ dimension physical space -- and the fact that different errors $\delta P$ can be used on all the PEPS tensors.

\subsection{Heralded errors}

The first set of errors that may occur in state preparation is the creation of a state that does not satisfy the dimer constraints. In other words, configurations such as
\begin{equation} \label{eq_heraldedmeasurementexample}
\begin{tikzpicture}[scale=0.4,baseline=-1,every node/.style={scale=0.8}]
\def\x{sqrt(3)}	
\draw[rounded corners] ({-0.75*\x}, -0.75) rectangle ({0.75*\x}, 0.75) {};
\draw[thick] ({0.5*\x},-0.5) -- ({-0.5*\x},0.5) -- ({-0.5*\x},-0.5) -- ({0.5*\x},0.5) -- ({0.5*\x},-0.5); 
\draw[PineGreen,fill=PineGreen] ({0.25*\x},0.25) circle (0.15); \draw[PineGreen,fill=PineGreen] ({-0.25*\x},-0.25) circle (0.15); 
\draw[BrickRed,thick] ({-0.5*\x}, 0.5) -- ({-0.5*\x}, 1); \draw[BrickRed,thick] ({0.5*\x}, 0.5) -- ({0.5*\x}, 1); \draw[BrickRed,thick] ({-0.5*\x}, -0.5) -- ({-0.5*\x}, -1); \draw[BrickRed,thick] ({0.5*\x}, -0.5) -- ({0.5*\x}, -1); 
\end{tikzpicture} , 
\end{equation}
have non-zero weight in the $\delta P$. Fortunately, this type of errors is heralded by fluorescence measurements. For the specific case of Eq.~\ref{eq_heraldedmeasurementexample} and one of the single qubit gate introduced in Secs.~\ref{ssec_wiredecoupling} and~\ref{ssec_singlequbitgate}, we would find $p_R$ and $p_L$ simultaneously non-zero. This case falling outside of the prescription Eq.~\ref{eq_PhysicalStatesDecouplingWires} heralds a failure of the dimer constraint and the results of the MBQC is know to be flawed. We may simply discard its result and run the algorithm again until no such error is observed. Breakdown of the dimer constraints different than that illustrated in Eq.~\ref{eq_heraldedmeasurementexample} can be diagnosed in a similar way.

This heralding scheme trades off deviation in the preparation of the dimer state with computation time, since the algorithm has to be run until the dimer constraint is satisfied throughout the lattice. For low densities of defects and low-depth quantum codes, this increase in computation time is not an issue. However, the rather high density of defects $d \simeq 20\%$ of Ref.~\cite{semeghini2021probing} only gives a 10\% success rate to 10 gates deep quantum codes, which is still too small to hope for the realization of a deep quantum code. Nevertheless, we are hopeful that, in the future, the quality of state preparation may be improved by engineering of the dynamical ramps of parameters~\cite{giudici2022dynamical}.

\subsection{Non-heralded gate errors}

Other errors $\delta P$ satisfying the dimer constraint cannot be heralded by fluorescence measurements, and directly impact the fidelity of the single and two qubit gates given in Sec.~\ref{sec_MBQCscheme} as $\mathcal{F} \simeq 1 - \eta p$ with $p$ is a constant. Without knowing the physical mechanisms giving the error $\delta P$, the best we can do is evaluate the coefficient $p$ by averaging over all possible errors that are not heralded by fluorescence, and estimate $\eta$ from experiments.

Let us first estimate the coefficient $p$. The non-heralded errors can be parameterized by eight matrices, one for each of the allowed physical configurations given in Eq.~\ref{eq_AllNonzeroCoefficients}. Upon applying one of the measurement schemes of Sec.~\ref{sec_MBQCscheme}, these matrices are combined into a unitary matrix $V$ according to Eq.~\ref{eq_PhysicalStatesDecouplingWires} or Eq.~\ref{eq_twoqubitgateprojection}. This results in a new orthogonal matrix, which we want to compare to the unitary $U = D_\varphi$ or $U = Q$ obtained without noise for the single qubit and two qubit case, respectively. Using $\mathcal{F} = \overline{\Tr \left[(U-\eta V)(U-\eta V)^\dagger\right]} / 4$ where the bar denotes an average over all unitary matrices $V$, we get $p \simeq \overline{\Tr\left[ (I-V)(I-V)^\dagger \right]}/4-1$ with $I$ the identity matrix. This simple average estimate gives $p=1$~\cite{mezzadri2007generate}.

Using this value, we can now try to infer the $\eta$ for non-heralded errors from the experimental results of Ref.~\cite{semeghini2021probing}. We will use their measure of the mean excitation number $\langle Z \rangle_{6L}$ along a closed loop containing six links of the Kagome lattice. For a perfect dimer, this expectation value should be equal to one, but $\langle Z \rangle_{6L} \simeq 0.14$ is measured. In our language, this is equal to the fidelity of six measurements including the error coming from defects $[\mathcal{F}/(1-d)]^6$, from which we infer $\eta \simeq 10\%$.

In the setup of Ref.~\cite{semeghini2021probing}, non-heralded and heralded errors both occurs approximately once every ten gates. While the heralded errors can be corrected by running the computation multiple times, the only way to be sure non-heralded measurement do not spoil the result of the MBQC calculations is through averaging. For a quantum code consisting of only ten gates, these post-selection and averaging requirements require to run the code 100-1000 time to reach reliable results.

\section{Conclusion}

Measurement-based quantum computation feeds from entangled many-body states, and can be realized if sufficient experimental control over each individual constituent of the may-body state exists. It is therefore particularly suited to arrays of Rydberg atoms, which have repeatedly demonstrated their strength as quantum simulators, and benefit from tried-and-tested laser manipulation techniques to measure all atoms individually.

In this article, we have shown that the entangled many-body state realized in Ref.~\cite{semeghini2021probing} together with the proposed measurement pulses enable universal measurement-based quantum computation. This establishes that Rydberg arrays could be used as a platform for universal quantum computing in the near future, and could potentially compete with superconducting quantum circuits due to the large number of available atoms (about 220 in Ref.~\cite{semeghini2021probing}).

The practical implementation of MBQC in Rydberg arrays will however require much engineering efforts to run deep quantum codes due to the heralded and non-heralded errors highlighted in Sec.~\ref{sec_imperfections}. For current experimental conditions, error rates inferred from measurements indicate that only quantum code consisting of about ten gates can be reliably implemented.

This calls for a search of robust and resilient Rydberg phases with identical computational power, over which single qubit operations could be performed with circuits of smaller depth. It is also important to explore how to identify, quantify and correct the intrinsic errors arising from the non-adiabatic state preparation in Rydberg arrays, and how to mitigate the effects of the long-range part of the van der Waals interaction through pulse engineering.

\section*{Acknowledgments} 

I gratefully acknowledges support from the MathWorks fellowship, and thank A. Wei and I. L. Chuang for valuable insights and comments. I am also grateful to K. Zimmerman, J. Gallagher and their team for their precious help during last editing phase of the manuscript. 

\section*{Data availability}

The data that support the findings of this study are available from the corresponding author upon reasonable request.

\section*{Author declarations}

\subsection*{Conflict of interest}

The authors have no conflicts to disclose.

\bibliography{BiblioTermPaperQuantInfo}

\appendix

\section{Pulses and projections}

\subsection{Pulse (I)}

In the basis $\{
\begin{tikzpicture}[scale=0.3,baseline=-2.7]
\def\x{sqrt(3)}; \draw[thick] (0,0.5) -- ({0.5*\x},0) -- (0,-0.5); 
\end{tikzpicture} , \begin{tikzpicture}[scale=0.3,baseline=-2.7]
\def\x{sqrt(3)}; \draw[thick] (0,0.5) -- ({0.5*\x},0) -- (0,-0.5);  \draw[PineGreen,fill=PineGreen] ({0.25*\x},0.25) circle (0.2); 
\end{tikzpicture} , \begin{tikzpicture}[scale=0.3,baseline=-2.7]
\def\x{sqrt(3)}; \draw[thick] (0,0.5) -- ({0.5*\x},0) -- (0,-0.5);  \draw[PineGreen,fill=PineGreen] ({0.25*\x},-0.25) circle (0.2); 
\end{tikzpicture} \}$ introduced in the main text, the blockaded Hamiltonian reads 
\begin{equation}
\mathcal{H} = \Delta \begin{bmatrix} 0 & x/2& x/2\\x^*/2&-1&0 \\ x^*/2&0&-1 \end{bmatrix} , \quad |x| = \sqrt{\frac{3}{2}} . 
\end{equation}
This Hamiltonian is exponentiated in Eq.~\ref{eq_pulse1_evolution} for $\tau = \pi/\Delta$ and yields the effective measurement of Eq.~\ref{eq_EffectiveMeasurement}. The eigenvectors of this delayed measurements are the new eigenstates obtained after fluroescence measurement. 

\subsection{Pulse (II)}

In the four dimensional blockaded basis $\{
\begin{tikzpicture}[scale=0.3,baseline=-2.7]
\def\x{sqrt(3)}; \draw[thick] ({0.5*\x},0) -- (0,0.5) -- (0,-0.5) -- ({0.5*\x},0); 
\end{tikzpicture} , \begin{tikzpicture}[scale=0.3,baseline=-2.7]
\def\x{sqrt(3)}; \draw[thick] ({0.5*\x},0) -- (0,0.5) -- (0,-0.5) -- ({0.5*\x},0); \draw[PineGreen,fill=PineGreen] (0,0) circle (0.2); 
\end{tikzpicture} , \begin{tikzpicture}[scale=0.3,baseline=-2.7]
\def\x{sqrt(3)}; \draw[thick] ({0.5*\x},0) -- (0,0.5) -- (0,-0.5) -- ({0.5*\x},0); \draw[PineGreen,fill=PineGreen] ({0.25*\x},0.25) circle (0.2); 
\end{tikzpicture} , \begin{tikzpicture}[scale=0.3,baseline=-2.7]
\def\x{sqrt(3)}; \draw[thick] ({0.5*\x},0) -- (0,0.5) -- (0,-0.5) -- ({0.5*\x},0); \draw[PineGreen,fill=PineGreen] ({0.25*\x},-0.25) circle (0.2); 
\end{tikzpicture} \}$ for pulse (II), the Hamiltonian Eq.~\ref{eq_totalHamiltonian} reads
\begin{equation}
\mathcal{H} = \frac{\Omega}{2} \begin{bmatrix} 0&i&i&i\\-i&0&0&0\\-i&0&0&0\\-i&0&0&0 \end{bmatrix} , 
\end{equation}
where we have assumed a purely imaginary Rabi frequency for simplicity~\footnote{The phase of $\Omega$ can always be tailored using pulse (III)~\cite{verresen2021prediction}}. The evolution under this Hamiltonian for a time $\tau = 4 \pi / (3\sqrt{3} \Omega)$ is then described by 
\begin{equation}
U = e^{-i\tau\mathcal{H}} = - \frac{1}{2} \begin{bmatrix}
1 & -1& -1& -1 \\ 1 &-1&1&1 \\ 1 &1&-1&1 \\  1&1&1&-1
\end{bmatrix} .
\end{equation}
The four states distinguished by fluorescence measurements at the end of this evolution correspond to projections of the system before the blockaded evolution onto
\begin{subequations}
\begin{eqnarray}
U^\dagger \ket{\begin{tikzpicture}[scale=0.3,baseline=-2.7]
\def\x{sqrt(3)}; \draw[thick] ({0.5*\x},0) -- (0,0.5) -- (0,-0.5) -- ({0.5*\x},0); 
\end{tikzpicture}} &=& \frac{1}{2} \left[ - \begin{tikzpicture}[scale=0.3,baseline=-2.7]
\def\x{sqrt(3)}; \draw[thick] ({0.5*\x},0) -- (0,0.5) -- (0,-0.5) -- ({0.5*\x},0); 
\end{tikzpicture} +\begin{tikzpicture}[scale=0.3,baseline=-2.7]
\def\x{sqrt(3)}; \draw[thick] ({0.5*\x},0) -- (0,0.5) -- (0,-0.5) -- ({0.5*\x},0); \draw[PineGreen,fill=PineGreen] (0,0) circle (0.2); 
\end{tikzpicture} +  \begin{tikzpicture}[scale=0.3,baseline=-2.7]
\def\x{sqrt(3)}; \draw[thick] ({0.5*\x},0) -- (0,0.5) -- (0,-0.5) -- ({0.5*\x},0); \draw[PineGreen,fill=PineGreen] ({0.25*\x},0.25) circle (0.2); 
\end{tikzpicture}+\begin{tikzpicture}[scale=0.3,baseline=-2.7]
\def\x{sqrt(3)}; \draw[thick] ({0.5*\x},0) -- (0,0.5) -- (0,-0.5) -- ({0.5*\x},0); \draw[PineGreen,fill=PineGreen] ({0.25*\x},-0.25) circle (0.2); 
\end{tikzpicture} \right] ,\\
U^\dagger \ket{\begin{tikzpicture}[scale=0.3,baseline=-2.7]
\def\x{sqrt(3)}; \draw[thick] ({0.5*\x},0) -- (0,0.5) -- (0,-0.5) -- ({0.5*\x},0); \draw[PineGreen,fill=PineGreen] (0,0) circle (0.2); 
\end{tikzpicture}} &=& \frac{1}{2} \left[ - \begin{tikzpicture}[scale=0.3,baseline=-2.7]
\def\x{sqrt(3)}; \draw[thick] ({0.5*\x},0) -- (0,0.5) -- (0,-0.5) -- ({0.5*\x},0); 
\end{tikzpicture} +\begin{tikzpicture}[scale=0.3,baseline=-2.7]
\def\x{sqrt(3)}; \draw[thick] ({0.5*\x},0) -- (0,0.5) -- (0,-0.5) -- ({0.5*\x},0); \draw[PineGreen,fill=PineGreen] (0,0) circle (0.2); 
\end{tikzpicture} - \begin{tikzpicture}[scale=0.3,baseline=-2.7]
\def\x{sqrt(3)}; \draw[thick] ({0.5*\x},0) -- (0,0.5) -- (0,-0.5) -- ({0.5*\x},0); \draw[PineGreen,fill=PineGreen] ({0.25*\x},0.25) circle (0.2); 
\end{tikzpicture} -\begin{tikzpicture}[scale=0.3,baseline=-2.7]
\def\x{sqrt(3)}; \draw[thick] ({0.5*\x},0) -- (0,0.5) -- (0,-0.5) -- ({0.5*\x},0); \draw[PineGreen,fill=PineGreen] ({0.25*\x},-0.25) circle (0.2); 
\end{tikzpicture}\right] ,\\
U^\dagger \ket{\begin{tikzpicture}[scale=0.3,baseline=-2.7]
\def\x{sqrt(3)}; \draw[thick] ({0.5*\x},0) -- (0,0.5) -- (0,-0.5) -- ({0.5*\x},0); \draw[PineGreen,fill=PineGreen] ({0.25*\x},0.25) circle (0.2); 
\end{tikzpicture}} &=& \frac{1}{2} \left[ - \begin{tikzpicture}[scale=0.3,baseline=-2.7]
\def\x{sqrt(3)}; \draw[thick] ({0.5*\x},0) -- (0,0.5) -- (0,-0.5) -- ({0.5*\x},0); 
\end{tikzpicture} -\begin{tikzpicture}[scale=0.3,baseline=-2.7]
\def\x{sqrt(3)}; \draw[thick] ({0.5*\x},0) -- (0,0.5) -- (0,-0.5) -- ({0.5*\x},0); \draw[PineGreen,fill=PineGreen] (0,0) circle (0.2); 
\end{tikzpicture} + \begin{tikzpicture}[scale=0.3,baseline=-2.7]
\def\x{sqrt(3)}; \draw[thick] ({0.5*\x},0) -- (0,0.5) -- (0,-0.5) -- ({0.5*\x},0); \draw[PineGreen,fill=PineGreen] ({0.25*\x},0.25) circle (0.2); 
\end{tikzpicture} -\begin{tikzpicture}[scale=0.3,baseline=-2.7]
\def\x{sqrt(3)}; \draw[thick] ({0.5*\x},0) -- (0,0.5) -- (0,-0.5) -- ({0.5*\x},0); \draw[PineGreen,fill=PineGreen] ({0.25*\x},-0.25) circle (0.2); 
\end{tikzpicture}\right] ,\\
U^\dagger \ket{\begin{tikzpicture}[scale=0.3,baseline=-2.7]
\def\x{sqrt(3)}; \draw[thick] ({0.5*\x},0) -- (0,0.5) -- (0,-0.5) -- ({0.5*\x},0); \draw[PineGreen,fill=PineGreen] ({0.25*\x},-0.25) circle (0.2); 
\end{tikzpicture}} &=& \frac{1}{2} \left[ - \begin{tikzpicture}[scale=0.3,baseline=-2.7]
\def\x{sqrt(3)}; \draw[thick] ({0.5*\x},0) -- (0,0.5) -- (0,-0.5) -- ({0.5*\x},0); 
\end{tikzpicture} -\begin{tikzpicture}[scale=0.3,baseline=-2.7]
\def\x{sqrt(3)}; \draw[thick] ({0.5*\x},0) -- (0,0.5) -- (0,-0.5) -- ({0.5*\x},0); \draw[PineGreen,fill=PineGreen] (0,0) circle (0.2); 
\end{tikzpicture} - \begin{tikzpicture}[scale=0.3,baseline=-2.7]
\def\x{sqrt(3)}; \draw[thick] ({0.5*\x},0) -- (0,0.5) -- (0,-0.5) -- ({0.5*\x},0); \draw[PineGreen,fill=PineGreen] ({0.25*\x},0.25) circle (0.2); 
\end{tikzpicture} +\begin{tikzpicture}[scale=0.3,baseline=-2.7]
\def\x{sqrt(3)}; \draw[thick] ({0.5*\x},0) -- (0,0.5) -- (0,-0.5) -- ({0.5*\x},0); \draw[PineGreen,fill=PineGreen] ({0.25*\x},-0.25) circle (0.2); 
\end{tikzpicture} \right] .
\end{eqnarray}
\end{subequations}
They precisely corresponds to the states $\frac{1}{2} (
\begin{tikzpicture}[scale=0.3,baseline=-2.7]
\def\x{sqrt(3)}; \draw[thick] ({0.5*\x},0) -- (0,0.5) -- (0,-0.5) -- ({0.5*\x},0); 
\end{tikzpicture} - p_\uparrow p_\downarrow \begin{tikzpicture}[scale=0.3,baseline=-2.7]
\def\x{sqrt(3)}; \draw[thick] ({0.5*\x},0) -- (0,0.5) -- (0,-0.5) -- ({0.5*\x},0); \draw[PineGreen,fill=PineGreen] (0,0) circle (0.2); 
\end{tikzpicture}  - x p_\uparrow \begin{tikzpicture}[scale=0.3,baseline=-2.7]
\def\x{sqrt(3)}; \draw[thick] ({0.5*\x},0) -- (0,0.5) -- (0,-0.5) -- ({0.5*\x},0); \draw[PineGreen,fill=PineGreen] ({0.25*\x},-0.25) circle (0.2); 
\end{tikzpicture}  - x p_\downarrow \begin{tikzpicture}[scale=0.3,baseline=-2.7]
\def\x{sqrt(3)}; \draw[thick] ({0.5*\x},0) -- (0,0.5) -- (0,-0.5) -- ({0.5*\x},0); \draw[PineGreen,fill=PineGreen] ({0.25*\x},0.25) circle (0.2); 
\end{tikzpicture} )$ given in the main text, provided $x$/$p_\downarrow$/$p_\uparrow$ are equal to one if the left/bottom right/top right link of the triangle is in $\ket{e}$, and to minus one otherwise.

\subsection{Pulse (III)}

The last pulse introduced is a standard dephasing operation.

\section{Gates} \label{app_gate}

\subsection{Wire decoupling}

The projection operator $D_0$ projects the physical space of the PEPS tensor on a superposition of six of its eight non-zero coefficients. For instance, if $p_R=0$, the projection is on the physical state
\begin{equation} \begin{split}
\psi & =
\begin{tikzpicture}[scale=0.3,baseline=-2.7] \def\x{sqrt(3)}; \draw[thick] ({\x},-0.5) -- (0,0.5) -- (0,-0.5) -- ({\x},0.5) -- ({\x},-0.5); \draw[PineGreen,fill=PineGreen] ({0.25*\x},0.25) circle (0.2); \end{tikzpicture} 
+ x_L \begin{tikzpicture}[scale=0.3,baseline=-2.7] \def\x{sqrt(3)}; \draw[thick] ({\x},-0.5) -- (0,0.5) -- (0,-0.5) -- ({\x},0.5) -- ({\x},-0.5); \draw[PineGreen,fill=PineGreen] ({0.25*\x},0.25) circle (0.2); \draw[PineGreen,fill=PineGreen] (0,0) circle (0.2); \end{tikzpicture} 
+ p_L \begin{tikzpicture}[scale=0.3,baseline=-2.7] \def\x{sqrt(3)}; \draw[thick] ({\x},-0.5) -- (0,0.5) -- (0,-0.5) -- ({\x},0.5) -- ({\x},-0.5); \draw[PineGreen,fill=PineGreen] ({0.25*\x},-0.25) circle (0.2); \end{tikzpicture} 
+ x_L p_L \begin{tikzpicture}[scale=0.3,baseline=-2.7] \def\x{sqrt(3)}; \draw[thick] ({\x},-0.5) -- (0,0.5) -- (0,-0.5) -- ({\x},0.5) -- ({\x},-0.5); \draw[PineGreen,fill=PineGreen] ({0.25*\x},-0.25) circle (0.2); \draw[PineGreen,fill=PineGreen] (0,0) circle (0.2); \end{tikzpicture} \\
& + x_R \left(
\begin{tikzpicture}[scale=0.3,baseline=-2.7] \def\x{sqrt(3)}; \draw[thick] ({\x},-0.5) -- (0,0.5) -- (0,-0.5) -- ({\x},0.5) -- ({\x},-0.5); \draw[PineGreen,fill=PineGreen] ({0.25*\x},0.25) circle (0.2); \draw[PineGreen,fill=PineGreen] ({\x},0) circle (0.2);  \end{tikzpicture} 
+ x_L \begin{tikzpicture}[scale=0.3,baseline=-2.7] \def\x{sqrt(3)}; \draw[thick] ({\x},-0.5) -- (0,0.5) -- (0,-0.5) -- ({\x},0.5) -- ({\x},-0.5); \draw[PineGreen,fill=PineGreen] ({0.25*\x},0.25) circle (0.2); \draw[PineGreen,fill=PineGreen] (0,0) circle (0.2); \draw[PineGreen,fill=PineGreen] ({\x},0) circle (0.2); \end{tikzpicture} 
+ p_L \begin{tikzpicture}[scale=0.3,baseline=-2.7] \def\x{sqrt(3)}; \draw[thick] ({\x},-0.5) -- (0,0.5) -- (0,-0.5) -- ({\x},0.5) -- ({\x},-0.5); \draw[PineGreen,fill=PineGreen] ({0.25*\x},-0.25) circle (0.2); \draw[PineGreen,fill=PineGreen] ({\x},0) circle (0.2); \end{tikzpicture} 
+ x_L p_L \begin{tikzpicture}[scale=0.3,baseline=-2.7] \def\x{sqrt(3)}; \draw[thick] ({\x},-0.5) -- (0,0.5) -- (0,-0.5) -- ({\x},0.5) -- ({\x},-0.5); \draw[PineGreen,fill=PineGreen] ({0.25*\x},-0.25) circle (0.2); \draw[PineGreen,fill=PineGreen] (0,0) circle (0.2); \draw[PineGreen,fill=PineGreen] ({\x},0) circle (0.2); \end{tikzpicture} \right) ,
\end{split} \end{equation}
as stated in the main text. Inferring the action of this projection on the virtual legs of the PEPS tensor using Eq.~\ref{eq_AllNonzeroCoefficients}, we see that the physically projected tensor is equivalent to a gate with the following action
\begin{subequations}
\begin{eqnarray}
\ket{1,+} &\mapsto& 1 \cdot \ket{-,1} , \\ 
\ket{1,-} &\mapsto& x_R \cdot \ket{-,0} , \\ 
\ket{0,+} &\mapsto& p_L \cdot \ket{+,1} , \\ 
\ket{0,-} &\mapsto& x_R p_L \cdot \ket{+,0} . 
\end{eqnarray}
\end{subequations}
The computation going from top to bottom, this can be rewritten as the following gate
\begin{equation} 
p_L \left[
\begin{tikzpicture}[scale=0.8,baseline=-2.5]
\draw[BrickRed,thick] (-1.2, 1) -- (-1.2, -1); \draw[fill=white] (-1.6, -0.9) rectangle (-0.8, -0.4) node[pos=.5] {$H$}; \draw[fill=white] (-0.8, 0.9) rectangle (-1.6, 0.4) node[pos=.5] {$Z^{\tilde{p}_L}$};
\draw[BrickRed,thick] (0.4, 1) -- (0.4, -1); \draw[fill=white] (0.8, 0.9) rectangle (0., 0.4) node[pos=.5] {$H$}; \draw[fill=white] (0.8, 0.25) rectangle (0., -0.25) node[pos=.5] {$Z^{\tilde{x}_R}$}; \draw[fill=white] (0.8, -0.9) rectangle (0., -0.4) node[pos=.5] {$X$};
\end{tikzpicture} \right] \, ,
\end{equation}
where we have defined tilde variables as in the main text $\tilde{q} = (1-q)/2$. Similarly, when $p_L=0$, the gate implemented is 
\begin{equation} 
p_R \left[
\begin{tikzpicture}[scale=0.8,baseline=-2.5]
\draw[BrickRed,thick] (-1.2, 1) -- (-1.2, -1); \draw[fill=white] (-0.8, 0.9) rectangle (-1.6, 0.4) node[pos=.5] {$Z^{\tilde{x}_L}$}; \draw[fill=white] (-1.6, -0.9) rectangle (-0.8, -0.4) node[pos=.5] {$H$}; \draw[fill=white] (-1.6, 0.25) rectangle (-0.8, -0.25) node[pos=.5] {$X$};
\draw[BrickRed,thick] (0.4, 1) -- (0.4, -1); \draw[fill=white] (0.8, 0.9) rectangle (0., 0.4) node[pos=.5] {$H$}; \draw[fill=white] (0.8, 0.25) rectangle (0., -0.25) node[pos=.5] {$Z^{\tilde{p}_R}$}; 
\end{tikzpicture} \right] \, .
\end{equation}
Up to a known global phase, those expression precisely agree with the one given in Eq.~\ref{eq_PhysicalStatesDecouplingWires} of the main text. 

\subsection{Single qubit operations}

When a pulse (III) is applied to the leftmost atoms of the bowtie before applying the wire decoupling scheme, only the $p_L=0$ part of our previous discussion changes. The $X$ appearing in the left computational wire is replaced by the gate
\begin{equation}
\begin{bmatrix} 0 & e^{-i \varphi /2} \\ e^{i \varphi/2} & 0 \end{bmatrix} ,
\end{equation}
which is precisely the gate $G_\varphi$ introduced in Eq.~\ref{eq_GatePtheta}.

\subsection{Two qubit gate}

The projection onto the physical states given in Eq.~\ref{eq_twoqubitgateprojection} yields an effective gate
\begin{equation} \label{appeq_q0} \begin{tikzpicture}[scale=0.8,baseline=-2.5]
\def\x{sqrt(3)}	; \draw[rounded corners] ({-0.5*\x}, -0.5) rectangle ({0.5*\x}, 0.5) {}; \node at (0,0) {$Q_0$} ;
\draw[BrickRed,thick] ({-0.3*\x}, -0.5) -- ({-0.3*\x}, -1.2); \draw[BrickRed,thick] ({0.3*\x}, -0.5) -- ({0.3*\x}, -1.2); \draw[BrickRed,thick] ({-0.3*\x}, 0.5) -- ({-0.3*\x}, 1.2); \draw[BrickRed,thick] ({0.3*\x}, 0.5) -- ({0.3*\x}, 1.2); \draw[fill=white] ({-0.3*\x-0.4}, -1.1) rectangle ({-0.3*\x+0.4}, -0.6) node[pos=.5] {$H$}; \draw[fill=white] ({0.3*\x-0.4}, 1.1) rectangle ({0.3*\x+0.4}, 0.6) node[pos=.5] {$H$};
\end{tikzpicture} \end{equation}
where $Q_0$ has the following action
\begin{subequations} \begin{eqnarray}
\ket{0,0} &\mapsto& x^L p_\uparrow^L \ket{0,1} - x^R p_\uparrow^R \ket{1,0} , \\
\ket{1,1} &\mapsto& p_\uparrow^R p_\downarrow^R p_\uparrow^L p_\downarrow^L \left[ x^R p_\uparrow^R \ket{0,1} + x^L p_\uparrow^L \ket{1,0} \right] , \\
\ket{0,1} &\mapsto& x^L p_\uparrow^L p_\uparrow^R p_\downarrow^R \ket{0,0} - x^R p_\downarrow^R \ket{1,1} , \\ 
\ket{1,0} &\mapsto& x^R p_\uparrow^R p_\uparrow^L p_\downarrow^L \ket{0,0} + x^L p_\downarrow^L \ket{1,1} .
\end{eqnarray} \end{subequations}
The various phases appearing are simple Pauli errors that can be removed if we rewrite
\begin{equation}
\begin{tikzpicture}[scale=0.8,baseline=-2.5]
\def\x{sqrt(3)}	; \draw[rounded corners] ({-0.5*\x}, -0.5) rectangle ({0.5*\x}, 0.5) {}; \node at (0,0) {$Q_0$} ;
\draw[BrickRed,thick] ({-0.3*\x}, -0.5) -- ({-0.3*\x}, -1.2); \draw[BrickRed,thick] ({0.3*\x}, -0.5) -- ({0.3*\x}, -1.2); \draw[BrickRed,thick] ({-0.3*\x}, 0.5) -- ({-0.3*\x}, 1.2); \draw[BrickRed,thick] ({0.3*\x}, 0.5) -- ({0.3*\x}, 1.2); 
\end{tikzpicture}  =
\begin{tikzpicture}[scale=0.8,baseline=-2.5]
\def\x{sqrt(3)}	; \draw[rounded corners] ({-0.5*\x}, -0.5) rectangle ({0.5*\x}, 0.5) {}; \node at (0,0) {$Q_1$} ;
\draw[BrickRed,thick] ({-0.3*\x}, -0.5) -- ({-0.3*\x}, -1.2); \draw[BrickRed,thick] ({0.3*\x}, -0.5) -- ({0.3*\x}, -1.2); \draw[BrickRed,thick] ({-0.3*\x}, 0.5) -- ({-0.3*\x}, 1.2); \draw[BrickRed,thick] ({0.3*\x}, 0.5) -- ({0.3*\x}, 1.2); \draw[fill=white] ({-0.3*\x-0.4}, 1.1) rectangle ({-0.3*\x+0.4}, 0.6) node[pos=.5] {$Z^{z_t^L}$}; \draw[fill=white] ({-0.3*\x-0.4}, -1.1) rectangle ({-0.3*\x+0.4}, -0.6) node[pos=.5] {$Z^{z_b^L}$}; \draw[fill=white] ({0.3*\x-0.4}, 1.1) rectangle ({0.3*\x+0.4}, 0.6) node[pos=.5] {$Z^{z_t^R}$}; \draw[fill=white] ({0.3*\x-0.4}, -1.1) rectangle ({0.3*\x+0.4}, -0.6) node[pos=.5] {$Z^{z_b^R}$};
\end{tikzpicture} ,
\end{equation}
with $z_b^W = \widetilde{x^W p_\uparrow^W}$ and $z_t^W = \widetilde{x^W p_\downarrow^W p_\uparrow^R p_\downarrow^L}$ for $W=R/L$, which gives a simple action of $Q_1$
\begin{subequations} \begin{eqnarray}
\ket{0,0} &\mapsto&  \ket{0,1} -  \ket{1,0} , \\
\ket{1,1} &\mapsto&  \ket{0,1} +  \ket{1,0}  , \\
\ket{0,1} &\mapsto& \ket{0,0} - \ket{1,1} , \\ 
\ket{1,0} &\mapsto& \ket{0,0} + \ket{1,1} .
\end{eqnarray} \end{subequations}
We may rewrite $Q_1$ as 
\begin{equation}
\begin{tikzpicture}[scale=0.8,baseline=-2.5]
\def\x{sqrt(3)}; \draw[BrickRed,thick] ({-0.3*\x}, 1) -- ({-0.3*\x}, -1); \draw[BrickRed,thick] ({0.3*\x}, 1) -- ({0.3*\x}, -1); 
\draw ({-0.3*\x}, -0.15) -- ({0.3*\x+0.15}, -0.15); \draw[fill=black] ({-0.3*\x},-0.15) circle (0.15); \draw[black] ({0.3*\x},-0.15) circle (0.15); 
\draw[fill=white] ({-0.3*\x-0.4}, 0.05) rectangle ({-0.3*\x+0.4}, 0.55) node[pos=.5] {$H$}; \draw[fill=white] ({0.3*\x-0.4}, 0.05) rectangle ({0.3*\x+0.4}, 0.55) node[pos=.5] {$X$}; \draw[fill=white] ({-0.3*\x-0.4}, -0.35) rectangle ({-0.3*\x+0.4}, -0.85) node[pos=.5] {$Z$};
\draw ({-0.3*\x}, 0.75) -- ({0.3*\x+0.15}, 0.75); \draw[fill=black] ({-0.3*\x},0.75) circle (0.15); \draw[black] ({0.3*\x},0.75) circle (0.15); 
\end{tikzpicture} ,
\end{equation}
which reduces to the operator given in Eq.~\ref{eq_twoqubitgateprojection} up to Pauli errors arising from the commutation of the bottom-left Hadamard gate of Eq.~\ref{appeq_q0} and the $Z^{z_t}$ operators with $Q_1$. 

\end{document}